\documentclass[5p]{elsarticle}

\usepackage{lineno,hyperref}

\usepackage{bm}
\usepackage{amsmath}
\newcommand{\mvector}[1]{{\bm #1}}

\usepackage{dontIndentFirst}
\usepackage{color}

\journal{Computer Physics Communications}

\begin{document}

\begin{frontmatter}

\title{\underline{Bio}t \underline{S}avart L\underline{aw} integrator BioSaw}
\cortext[mycorrespondingauthor]{Corresponding author}
\author[aalto]{Simppa {\"Ak\"aslompolo}\corref{mycorrespondingauthor}}

\ead{simppa.akaslompolo@alumni.aalto.fi}
\address[aalto]{Department of Applied Physics, Aalto University, FI-00076 AALTO, FINLAND}
\author[aalto]{Tuomas Koskela}
\author[aalto]{Taina Kurki-Suonio}

\begin{abstract}
This contribution documents the methods used in the BioSaw code. The code is inteded to be a flexible tool for calculating magnetic fields due to coils in magnetic confinement fusion devices. It assumes the conductors are infinitesimally thin and can be described as either point sequences or circular coils. The code can calculate both the magnetic field as well as the vector potential due to the coils. The fields can be reduced very near the coils to avoid singular behaviour caused by the thin conductor approximation.
\end{abstract}


\end{frontmatter}

\section{Introduction}
Magnetic fields play central roles in many electric devices, such as motors, magnetic resonance imaging devices and speakers.
A particular case motivating this work is magnetic confinement fusion where the plasma is controlled purely with magnetic fields.
The \underline{Bio}t-\underline{S}avart \underline{law} integrator BioSaw was written to calculate the magnetic fields of the coils in tokamaks, but is generally applicable to any coil. 
Its support for both Cartesian and cylindrical output grids stems from the tokamak background. 

The resulting fields are typically used as part of the input of the \textsc{ascot} code~\cite{ascot4ref}.
The closest known relatives of BioSaw are the \textsc{vacfield} code~\cite{strumbergervacfield} and the \textsc{Biot Savart magnetic Toolbox}~\cite{BSmag}.
BioSaw is designed for high performance computing: it is quite fast and well parallelised (hybrid OpenMP+MPI).

BioSaw can calculate the magnetic field $\mvector{B}$ and vector potential $\mvector{A}$ due to an infinitesimally thin, stationary, static current carrying wire. 
The Biot-Savart law for the magnetic field and the vector potential describes the field at location $\mvector{r}$ due to current $I_0$ in a thin conductor. It is an integral along the coil:
\begin{eqnarray}
  \mvector{B}(\mvector{r}) &=& \frac{\mu_0I_0}{4\pi}\int\displaylimits_\mathrm{coil}\frac{\mathrm{d}\mvector{\ell}\times(\mvector{r}-\mvector{r}')}{|\mvector{r}-\mvector{r}'|^3}\label{eq:Bspline}\\
  \mvector{A}(\mvector{r}) &=& \frac{\mu_0I_0}{4\pi}\int\displaylimits_\mathrm{coil}\frac{\mathrm{d}\mvector{\ell}}{|\mvector{r}-\mvector{r}'|}\ ,\label{eq:Aspline}
\end{eqnarray}
 where $\mathrm{d}\mvector{\ell}$ denotes the coil element at location $\mvector{r}'$. 
The magnetic field (or magnetic flux density, $\mvector{B}$) is the curl of the magnetic vector potential: $\nabla\times \mvector{A}=\mvector{B}$.

BioSaw supports three different calculation methods for different kinds of coils:
Usually the code takes as input an ordered set of points on the coil.
An irregularly shaped but smooth, sparsely sampled coil can be represented by a spline curve. 
The second method is to sum the field over linear current segments (also called vortex lines in hydrodynamic applications) between the points.
The third case are circular coils: their fields are calculated from coil locations and radii using an analytic formula. 

This contribution has the following structure: sections \ref{sec:splines}, \ref{sec:vortexes} and \ref{sec:circulars} describe the three calculation methods.
Section \ref{sec:tapering} describes how the field strength can be tapered down very near the coils to avoid numerical problems. 
Section \ref{sec:validity} analyses the error due to approximating a macroscopic conductor as an infinitesimally thin current filament.
Section \ref{sec:verification} shows verification results of the code.
The contribution is concluded with a summary in section \ref{sec:summary}.

\section{Smooth coil interpolated with splines}
\label{sec:splines}
Equations~\eqref{eq:Bspline} and~\eqref{eq:Aspline} can be integrated directly using a numerical quadrature (\textsc{quadpack}\cite{quadpack} in BioSaw), if there is a way to evaluate the coil coordinates $\mvector{r}(s)$ and direction $\mathrm{d}\mvector{\ell}(s)$ at arbitrary coil location. Here $s$ is a parameter along the coil. This is achieved by fitting a spline~\cite{pspline} to the coil coordinates. A separate spline for each Cartesian coordinate $x$, $y$ and $z$ is used. The direction vector is simply the unit tangent vector of the splines. The spline approach works best for a smooth coil described by a relatively sparse sequence of points. For coils forming a closed loop, periodic boundary conditions can be used for the spline fitting.

Biosaw calculates the spline that passes through the point sequence. The resulting curve is a smooth third-order piece-wise polynomial. In some cases, especially with regular straight segments, the spline may start to oscillate between the points, ``taking detours'' between the points. The code tries to automatically detect such behaviour.  It is accomplished by comparing the lengths of the spline and the broken line presentation of the coil. If the length of the spline is more than 1\,\% longer than the broken line presentation, the program gives a warning.

\section{Coil from straight segments: vortexes}
\label{sec:vortexes}
If the coil can be best represented by a large number of short straight segments, the total field can be obtained by summing up contributions of these short segments. 
The field due to each segment is given by an analytic formula. 
For using the formula, the evaluation location must be first expressed in the coordinate system of the segment:
the straight thin conductor has one end situated at the origin and the other end at the $z$ axis, at $z=L$.
The field is evaluated at distance $R$ from the $z$ axis and at  plane $z=h$.
The magnetic field rotates around the segment and vector potential is parallel to the segment.

The formula are obtained by integrating equations~\eqref{eq:Bspline} and~\eqref{eq:Aspline} for the the straight segment. 
The cross product $\mathrm{d}\mvector{\ell}\times(\mvector{r}-\mvector{r}')$ in~\eqref{eq:Bspline}  reduces to $-R\textrm{d}s$. 
The following formulas are obtained with some calculus:

\begin{eqnarray}
  \label{eq:Bsegments}
  \mvector{B}(L,R,h)&=&\frac{\mu_0 I_0}{4\pi}\int_0^L -\frac{R\textrm{d}s}{\sqrt{R^2+(h-s)^2}^3}\ \hat{\mvector{\phi}}\\  
  &=&-\frac{\mu_0 I_0}{4\pi R} \left( \frac{L-h}{d_L}+\frac{h}{d_o}  \right)\ \hat{\mvector{\phi}}\label{eq:BsegmentsR}\\
  \label{eq:Asegments}
\mvector{A}(L,R,h)&=&\frac{\mu_0 I_0}{4\pi}\int_0^L \frac{\textrm{d}s}{\sqrt{R^2+(h-s)^2}}\ \hat{\mvector{z}}\\
&=&\frac{\mu_0 I_0}{4\pi }\ln\left(\frac{d_o           +h}{d_L               +L-h}\right)\ \hat{\mvector{z}}\ ,\label{eq:AsegmentsR}
\end{eqnarray}
where $\hat{\mvector{\phi}}$ and $\hat{\mvector{z}}$ are unit vectors of the cylindrical coordinates at the evaluation location, $d_o=\sqrt{R^2+h^2}$ and $d_L=\sqrt{R^2+(h-L)^2}$ are distances from the end points of the segment.
The fields from the formulas are transformed back to the global coordinate system before summing up the segments.

\section{Circular coils}
\label{sec:circulars}
The code can calculate the field due to a thin circular coil using an analytic formula~\cite{kunst2007currentLoop,ElectromagneticFieldBook}. 
Evaluating it is much faster and probably more accurate than evaluating the field using numerical integrals. 
The coil is described by the following parameters:  centre location, normal vector of the coil plane, radius ($a$), and the current in the coil ($I_0$). 

The formula is defined in the coordinate system of the coil: the origin is at the coil centre and the $xy$ plane is the coil plane. 
The field is evaluated at distance $R$ from the $z$ axis at plane $z=h$.
The formulas read:
\begin{align}
  \label{eq:Acirc}
  \mvector{B} &=& \frac{\mu_0I_0}{4\pi}\frac{k}{\sqrt{aR^3}} &\left[-h\left(K(k)-\frac{2-k^2}{2(1-k^2)}E(k)\right)\hat{\mvector{R}} +\cdots\right. \nonumber\\ 
&&& \ \ \left.R\left(K(k)+\frac{k^2(R+a)-2R}{2R(1-k^2)}E(k) \right)\hat{\mvector{z}} \right]\\
  \mvector{A} &=& \frac{\mu_0I_0}{4\pi}2k\sqrt{\frac{a}{R}} &\left[\left(\frac{2}{k}-k\right)K(k)-\frac{2}{k}E(k) \right]\hat{\mvector{\phi}}\ ,
\end{align} where $\hat{\mvector{R}}$, $\hat{\mvector{\phi}}$ and $\hat{\mvector{z}}$ are unit vectors of the cylindrical coordinates at the evaluation location and $k=\sqrt{\frac{4aR}{(R+a)^2+h^2}}$. 
After evaluating the fields, they are transformed back to the global coordinate system.

The formulas contain the complete elliptic integrals of the first and second kind,
\begin{eqnarray}
  \label{eq:ellipticIntegral}
  K(k)&=&\int_0^{2\pi}\frac{\mathrm{d}\alpha}{\sqrt{1-k^2\sin^2\alpha}}\\
  E(k)&=&\int_0^{2\pi}\sqrt{1-k^2\sin^2\alpha}\ \mathrm{d}\alpha\ ,
\end{eqnarray}
which are evaluated using the  \textsc{slatec} library~\cite{slatec} routines \textsc{drf} and \textsc{drd}.

\section{Tapering off the field near the coils}
\label{sec:tapering}
BioSaw assumes that the current conductors are infinitely thin. This is of course a good approximation only far from the conductor. 
The field behaves as $B\sim1/\rho^2$ and $A\sim 1/\rho$, where $\rho$ is distance from the coil. 
Often the field is interesting only sufficiently far from the coils, but the calculation grid still extends to the coils. 
In this case some of the evaluation points may reside near or on top of the coils so that $\rho$ approaches zero. 
The outcome would be very large values for the fields. 

To avoid such numerical problems, BioSaw includes an option to taper off the field down inside a user configurable effective coil radius $\rho_0$.
This is based on the behaviour of the  magnetic field in a current carrying wire with circular cross section: the field goes to zero at the centre.
Thus,  the leading diverging term $B\sim \rho^{-2}$ is simply cancelled by multiplying with $p_B(\rho)=\frac{\rho^2}{\rho^2_0}$. 
For the vector potential, the goal is to have a smooth transition at the effective radius and smooth behaviour at the coil centre. 
The potential value inside $\rho_0$ is multiplied with $p_A(\rho)$, which is defined by requiring that $p_A'(0)=0$, $p_A(1)=1$ and $p_A'(1)=\frac{d}{d\rho}A=\frac{d}{d\rho}\frac{1}{\rho}$, where the prime denotes the derivative. 

\section{On the validity of thin filament approximation}
\label{sec:validity}
\begin{figure}
  \centering
\includegraphics[width=0.3\textwidth]{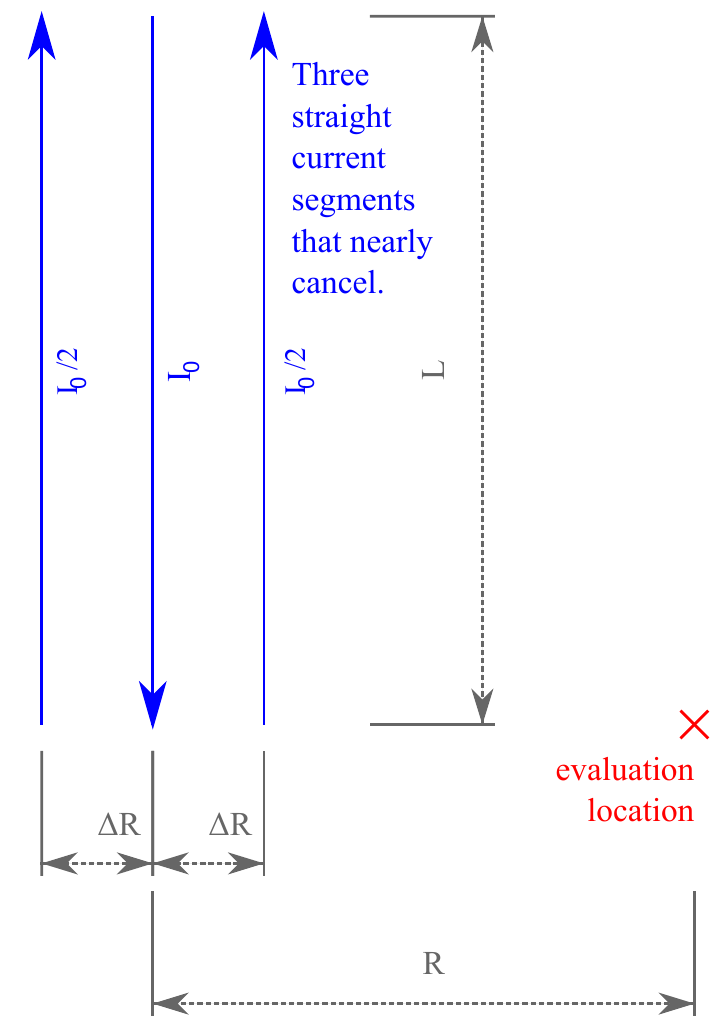}
  
  \caption{The geometry used for assessing when a thin conductor is a good approximation of a thick conductor.}
\label{fig:thicknessEffect}
\end{figure}

It is important to assess when a thick conductor or even a full coil can be represented by a single thin filament. 
This is studied by a modeling a coil carrying the current $I_0$ and consisting of a single straight current segment with length $L$.  
A thin conductor is placed in the middle of the coil.
The field produced by the thin coil is almost exactly canceled by a conductor with thickness $2\Delta R$ with the current flowing in the opposite direction.
To exacerbate the situation, all the current in the thick conductor is condensed into two thin current filaments located at the nearest and furthest parts of the conductor, as illustrated by figure \ref{fig:thicknessEffect}. 

The field is evaluated at distance $R$ from one end of the thin conductor.
An expression for the field due to the three current segments can be produced using equation \eqref{eq:BsegmentsR}.
The expression is then divided by the field due to the central conductor, to arrive at an expression for the \emph{relative error of the field due to thin conductor approximation}:
\begin{align}
1-&\frac{Rd_0}{2\sqrt{L^2+(R-\Delta R)^2} (R-\Delta R)}&\cdots\nonumber\\
-&\frac{Rd_0}{2\sqrt{L^2+(R+\Delta R)^2} (R+\Delta R)}&
\end{align}
A Taylor expansion,
\begin{equation}
\frac{\Delta R}{R}-\frac{4R^4+3L^2R^2}{2R^4+4L^2R^2+2L^4}\left(\frac{\Delta R}{R}\right)^2+\mathcal{O}\left(\frac{\Delta R}{R}\right)^3
\end{equation}
 demonstrates how the error vanishes only linearly:  $\sim\frac{\Delta R}{R}$. 
The same procedure performed for equation \eqref{eq:AsegmentsR} shows that the vector potential vanishes quadratically: $\sim\left(\frac{\Delta R}{R}\right)^2$.
The conclusion is that the conductor thickness is significant at the vicinity of the conductor: distances comparable to the conductor thickness.

\section{Verification}
\label{sec:verification}
In principle, the verification of BioSaw is easy, since there are many other codes and analytical solutions to the problem. 
For example, BioSaw, \textsc{vacfield}~\cite{strumbergervacfield} and  \textsc{Biot Savart magnetic Toolbox}~\cite{BSmag} produce a very similar  magnetic field for an ASDEX Upgrade in vessel coil~\cite{Suttrop2009290}. 
Quite surprisingly, a three-way-comparison between the codes shows differences of several percent.
However, further study of the discrepancy or further benchmark between the codes is beyond the scope of this contribution. 

\begin{figure*}
  \centering
  \includegraphics[width=0.85\textwidth,trim=1.9cm 1.55cm 2cm 1.25cm,clip=true]{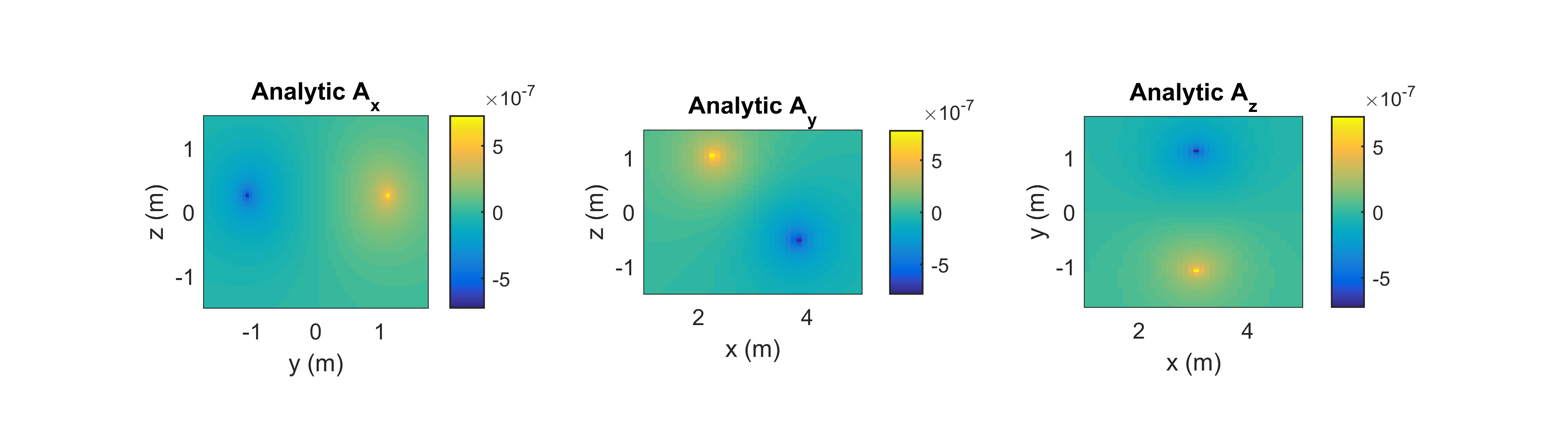}
  \includegraphics[width=0.85\textwidth,trim=1.9cm 1.60cm 2cm 1.25cm,clip=true]{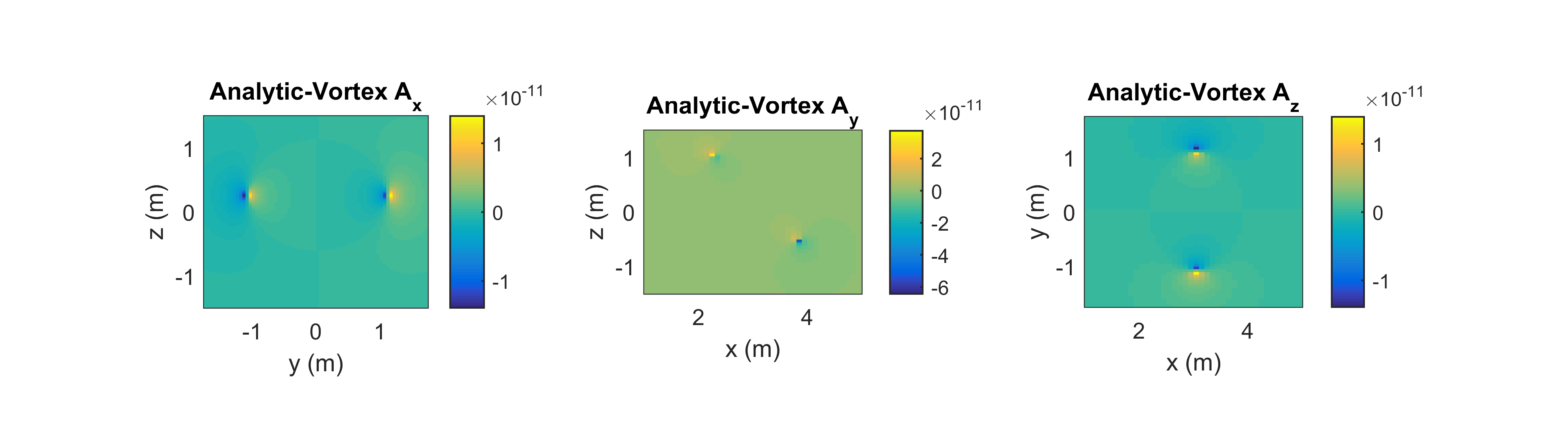}
  \includegraphics[width=0.85\textwidth,trim=1.9cm 1.586cm 2cm 1.25cm,clip=true]{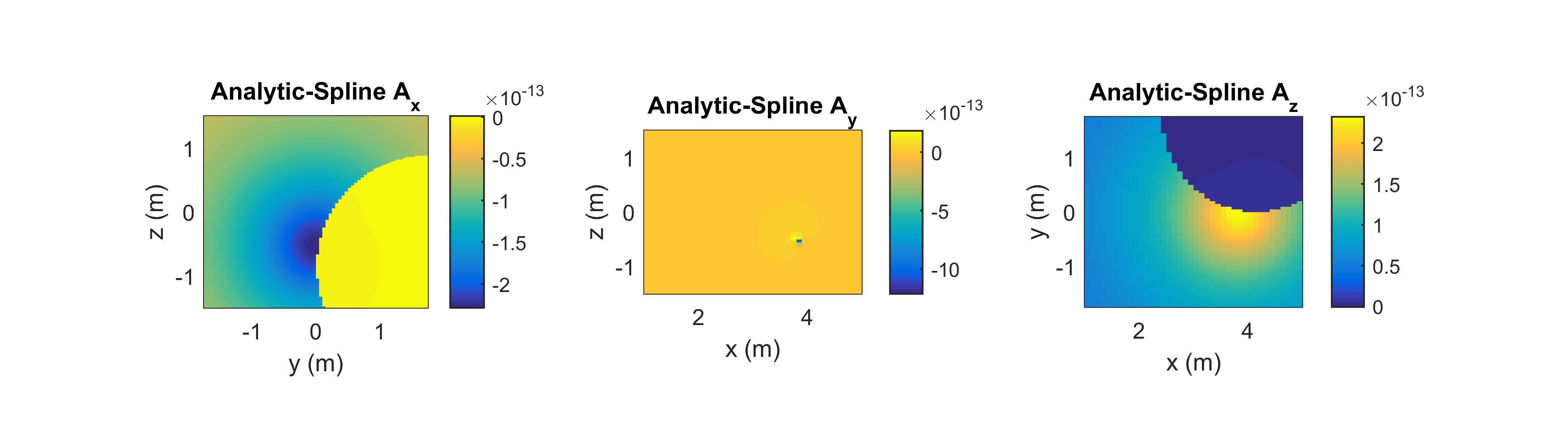}
  \includegraphics[width=0.85\textwidth,trim=1.9cm 1.55cm 2cm 1.25cm,clip=true]{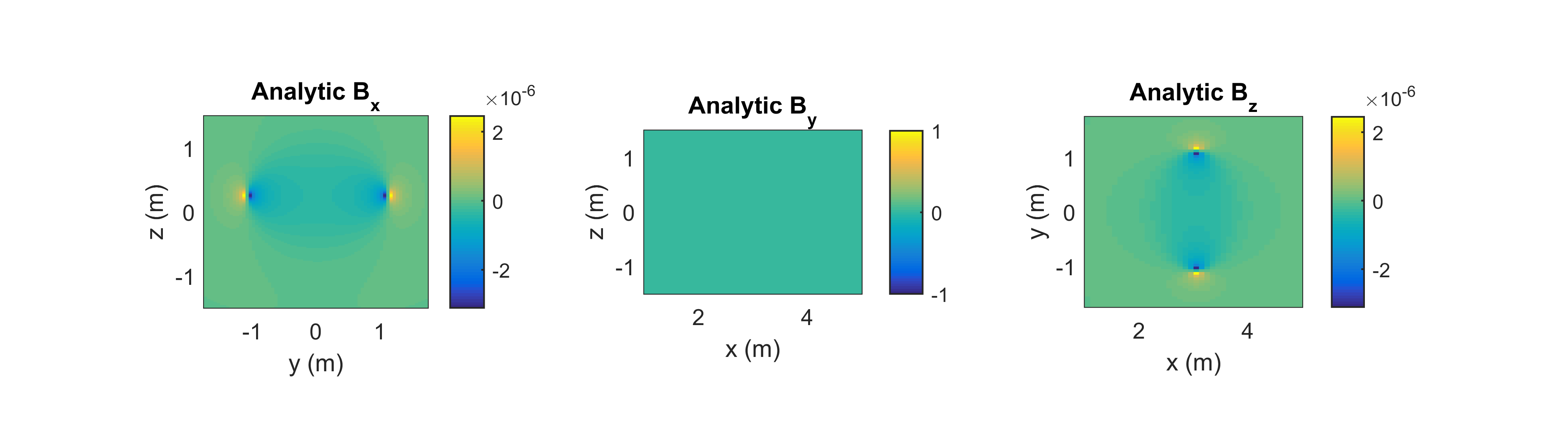}
  \includegraphics[width=0.85\textwidth,trim=1.9cm 1.60cm 2cm 1.25cm,clip=true]{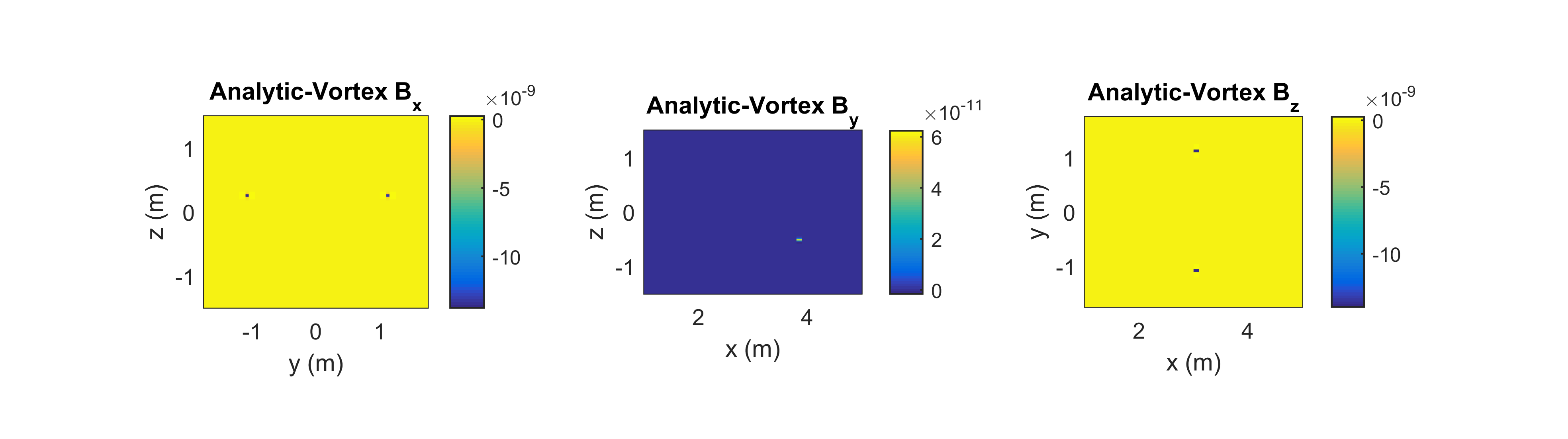}

  \includegraphics[width=0.85\textwidth,trim=1.9cm 0       2cm 1.25cm,clip=true]{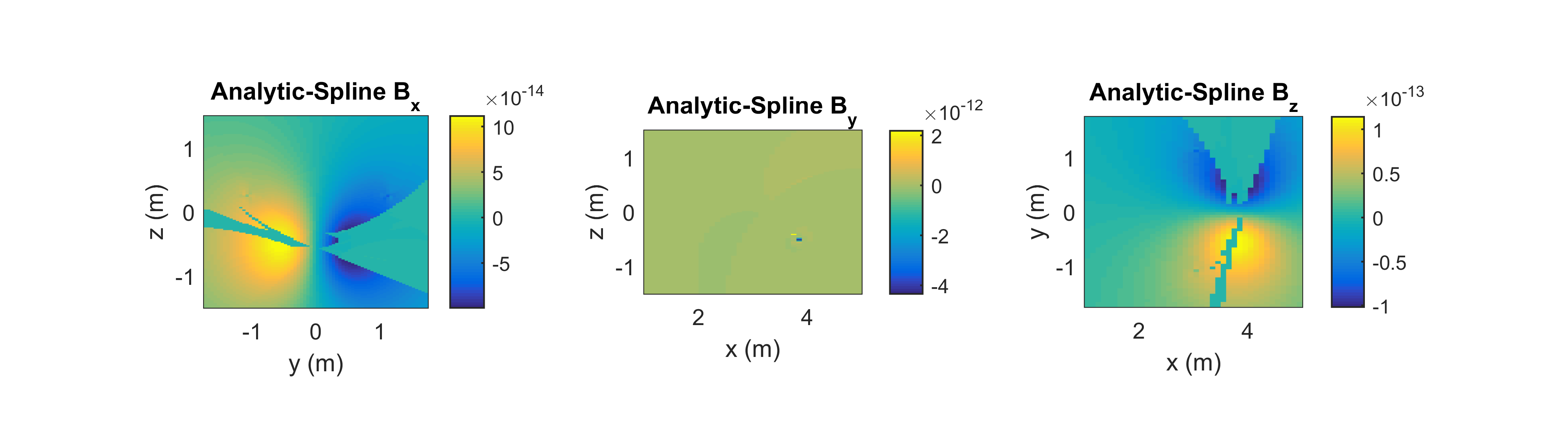}
  \caption{Comparison of the three implemented field calculation methods. The same field has been calculated using an analytic formula, spline integration and with linear segments (vortexes). The first and third rows are the field using analytic formula for a current loop. The other rows show difference to the analytic formula when using either the spline or the vortex integration method with 876 vortexes. The panels show the three orthogonal components of the magnetic vector potential $A$ in Vs/m and magnetic flux density $B$ in Teslas: the left column shows the $x$ component on the plane $x=3.00$\,m, the middle column shows the $y$ component on the plane $y=0.00$\,m and right column shows the $z$-component on the plane $z=0.24$\,m. The coil has radius 1.1\,m and nominal 1\,A current. The loop is tilted 45$^\circ$ around $Y$-axis and the loop is centered at $(x,y,z)$=$(3.00,0.00,0.25)$\,m.   }
  \label{fig:coils}
\end{figure*}

Comparison of the various integration methods implemented within BioSaw demonstrates the correct operation of the BioSaw code. The test case consists of a single circular coil with a nominal 1\,A current. The magnetic field and the vector potential are calculated with analytic formulas as well as with both of the integration methods.  Since the results are nearly identical, only the difference to the analytic solution is shown for the integrated fields (figure~\ref{fig:coils}). 

The precision of the analytic formula is best demonstrated by the $B_y$ field. The analytic formula properly reproduces the identically zero value (the coil is perpendicular to the $y$-plane). In both cases the numerical integrator methods produce fields that are correct to several decimal places. The Vortex integrator results are generally less precise for this coil than the spline version. The  spline interpolation has user definable tolerances. They were set to $10^{-10}$ for absolute error and for $10^{-5}$ for relative error, which seems to be achieved. The adaptive nature of the spline integration results in peculiar shapes for the errors in the illustrations.

\section{Summary}
\label{sec:summary}
This contribution described the main features and numerical methods of the BioSaw code.
It can be used to calculate the magnetic field and vector potential of arbitrary coils, as long as they can be described as a set of thin conductors. 

The error due to approximating macroscopic conductors or coils as thin current filaments was analysed.
The error in the magnetic field was found to diminish linearly with distance from the coil and the error in vector potential was found to diminish quadratically. 
Very near the thin conductor the field diverges numerically, which can be corrected by tapering off the field within a given distance from the coil.

The code was verified by calculating the field due to a single circular loop.
Both the magnetic field and the vector potential for the loop were calculated using the three different calculation methods implemented in the code.
The difference was found to be typically less than one part in ten thousand.

A possible future undertaking would be to do a careful benchmark between different Biot--Savart law integrator codes.

\section*{Acknowledgements}
This work was partially funded by the Academy of Finland project No. 259675. This work has received funding from Tekes - the Finnish Funding Agency for Innovation under the FinnFusion Consortium. The calculations presented above were performed using computer resources within the Aalto University School of Science "Science-IT" project.


\bibliography{../bibfile}

\end{document}